\documentclass[
 reprint,
 amsmath,amssymb,
 aps,
 longbibliography,
  pra,
 citeautoscript
]{revtex4-1}

\usepackage{graphicx}% Include figure files
\usepackage{dcolumn}% Align table columns on decimal point
\usepackage{bm}% bold math

\usepackage{color}

\newcommand{\bra}[1]{{\left\langle{#1}\right\vert}}
\newcommand{\ket}[1]{{\left\vert{#1}\right\rangle}}
\newcommand{\expec}[1]{{\langle{#1}\rangle}}

\def\FCW{1.0\columnwidth}

\begin{document}

\title{Quantum Data Compression of a Qubit Ensemble}

\author{Lee A. Rozema$^{1*}$}
 
\author{Dylan H. Mahler$^1$}

\author{Alex Hayat$^{1,2,3}$}

\author{Peter S. Turner$^{4\dagger}$}
 
\author{Aephraim M. Steinberg$^{1,3}$}

\affiliation{
$^1$Centre for Quantum Information \& Quantum Control and 
Dept. of Physics, University of Toronto, 60 St. George St., Toronto, Ontario, Canada M5S 1A7\\
$^2$Department of Electrical Engineering, Technion, Haifa 32000, Israel\\
$^3$Canadian Institute for Advanced Research, Toronto, Ontario M5G 1Z8, Canada\\
$^4$Department of Physics, Graduate School of Science, University of Tokyo, 7-3-1 Hongo, Bunkyo-ku, Tokyo, Japan 113-0033
}
 
\date{\today}
\begin{abstract}
Data compression is a ubiquitous aspect of modern information technology, and the advent of quantum information raises the question of what types of compression are feasible for quantum data, where it is especially relevant given the extreme difficulty involved in creating reliable quantum memories.  
We present a protocol in which an ensemble of quantum bits (qubits) can in principle be \textit{perfectly} compressed into exponentially fewer qubits.
We then experimentally implement our algorithm, compressing three photonic qubits into two.
{This protocol sheds light on the subtle differences between quantum and classical information.
 Furthermore, since data compression stores all of the available information about the quantum state in fewer physical qubits, it could provide a vast reduction in the amount of quantum memory required to store a quantum ensemble, making even today's limited quantum memories far more powerful than previously recognized.}
\end{abstract}
\maketitle

The amount of information that can be extracted from a classical system is precisely the same as the amount of information required for a complete description of the system's state. 
The same is not true quantum mechanically; to fully describe the state of a single quantum bit (qubit) would require an infinite amount of information, although no more than one (classical) bit of information can ever be extracted from a measurement of its quantum state.
Such fundamental differences between quantum and classical mechanics open up the possibility of new kinds of data compression with no classical analogue. 
In quantum mechanics an ensemble of identically prepared quantum systems provides much more information than a single copy -- this is not the case classically, where the information encoded in a single system's state can be accessed repeatedly.
Although quantum mechanically we cannot compress all of the information contained in an ensemble of systems down to a single quantum copy, we can achieve an exponential savings.
In this paper, we show how this exponential savings can be achieved using the  quantum Schur-Weyl transform \cite{bacon_efficient_2006,plesch_efficient_2010}, which can compress an ensemble of N identically prepared qubits into a memory of size $\log_2[N+1]$ qubits.
We show how the protocol can be made practical in an optical setting, experimentally implementing a three-qubit quantum circuit to compress a three-qubit ensemble into the state of two qubits.
To characterize this circuit, we show that we can perform measurements on the two compressed qubits, and still extract as much information as we would have been able to given all three original qubits.
Given our ability to extract information about measurements in multiple bases, we can conclude that the compressed state faithfully encodes the ``quantum information content'' of the original ensemble.
{Our results demonstrate that quantum memories can be used to store exponentially more information about a quantum state than would normally be expected for the number of physical qubits that the memory can hold.}

{From the point of view of estimation theory, a quantum state is never fully knowable, just as a classical probability distribution is not (both requiring an infinite amount of resources to be completely known).}
Hence, for our purposes, a quantum state is best thought of as a mathematical object which allows one to make \textit{testable} predictions about the statistics of potential measurements done on a large ensemble of identically prepared systems\footnote{This is not to say that the quantum state is not a real ontological object. Whether quantum states are real or  simply a description of our lack of knowledge, they are impossible to fully verify.  Just as it is impossible to fully verify a classical probability distribution.}.  
The task colloquially referred to as ``quantum state estimation'' is really the task of making possible predictions about the {\it expectation values} for observables which might be measured in the future.
Consider for instance estimating the spin projection of a qubit along a particular direction, given a fixed number of identically prepared copies of the qubit.
To do this the best strategy is quite simply to measure the the spin along the direction of interest on each copy and draw conclusions as one would do classically. 
Since quantum measurements are intrinsically uncertain and the state of each qubit collapses after measurement, having more copies allows one to make a better estimate.
If one does not know in advance which measurement will be of interest, the standard approach -- known as quantum state tomography -- is to reconstruct a \textit{density matrix} \cite{james_measurement_2001}, which contains enough information to allow one to estimate the expectation value of the spin along any direction.
This approach has the disadvantage that no single estimate can ever make optimal use of all of the available information \cite{blume-kohout_optimal_2010}.  
For instance, in single-qubit quantum state tomography one most commonly splits an initial ensemble of identical qubits into three equally sized groups, and measures $\hat{X}$ on all the members of one group, $\hat{Y}$ on another, and $\hat{Z}$ on the remaining group.  
But if, for example, one later wishes to estimate $\expec{\hat{Z}}$ (the expectation value of the spin along $\hat{Z}$), the measurements of $\hat{X}$ and $\hat{Y}$ (both of which are orthogonal to $\hat{Z}$) give no useful information, and $^2/_3$ of the measurements have been wasted.  
In fact, on average, whatever the projection of interest turns out to be, the estimate will be only as accurate as if that projection had been carried out on about one third of the ensemble; this is the price one pays for the generality of tomography: one has information about all three axes, but only one third as much information about each (the situation becomes more dire in higher dimensions, of course).
A better estimate of the spin along any specific direction could be made if one simply held onto all the copies of the system -- which would require a quantum memory -- until one knew which measurement was of interest, and then made this measurement on every single copy.
Thus, storing all of the qubits in a quantum memory would enable one to make much more accurate predictions about any single measurement than performing quantum state tomography.  
Note that in the classical case, an identically prepared ensemble of bits is highly redundant, so that ideally the information can be compressed down to just one bit.  
This redundancy, along with the daunting challenge of building large, high-fidelity quantum memories, motivates the question: how many qubits must we store to achieve the same prediction accuracy that is possible with the initial ensemble?

\begin{figure}%[p!]
\includegraphics[width=\FCW]{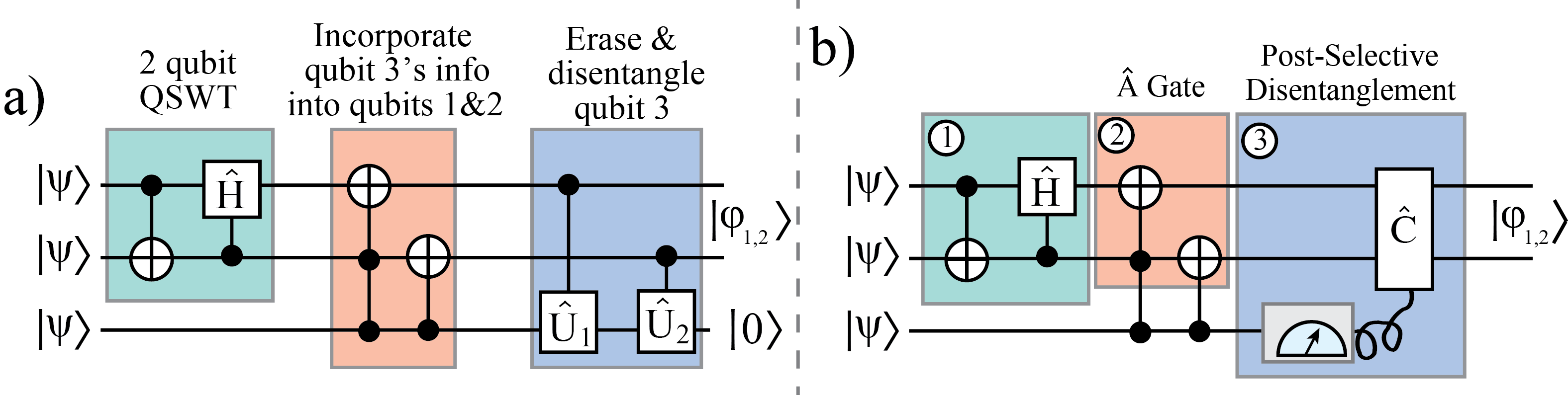}
\caption{\label{fig:1} 
{\bf a) Quantum Schur-Weyl Transform ---} A three-qubit implementation of the full quantum Schur-Weyl transform. 
{\textbf{$\hat{U}_1$} and \textbf{$\hat{U}_2$} are unitaries (whose detailed descriptions can be found in the main text) which are controlled by the upper qubit; the H is a controlled
Hadamard and the other two-qubit gates are CNOTs, while the three-qubit gate is a Toffoli.}
{\bf b) Simplified circuit ---} If the three input states are guaranteed to be identical pure states, then the final two controlled unitary gates can be replaced by a measurement-and-feed-forward scheme.
The shaded area labelled 2 can be viewed as a two-qubit unitary gate, $\hat{A}$, acting on the first two qubits which is controlled by the third qubit. 
\textbf{C} is a two-qubit unitary gate which is applied (or not) based on a measurement of qubit 3.
The numbered boxes correspond to the areas in figure 2 which show the physical implementation of the circuit elements.
}
\end{figure}

The dimension of the Hilbert space of an $N$-qubit system grows exponentially in the number of qubits, that is, as $2^N$. 
However, the state of an ensemble of N identically prepared (pure) qubits is the tensor product of N identical pure states, and lives in the $(N+1)$-dimensional fully symmetric subspace.
{Such N-qubit states can be described using $N+1$ rather than $2^N$ dimensions because the vast majority of the information in a general multi-qubit state describes the permutations of the qubits, which are irrelevant for an ensemble of identical qubits.
The remainder of the information describes angular momentum of the multi-qubit state, and is the only information relevant to estimating expectation values of single-qubit observables.}
Thus it is natural to ask if the initial $N$-qubit ensemble can be mapped reversibly (unitarily) onto exponentially fewer ($\log_2[N+1]$) qubits. 
In fact, this mapping of the computational basis into a new basis, separating the permutation from the angular momentum information, is well understood as the quantum Schur-Weyl transform (QSWT) \cite{bacon_efficient_2006}, and has been theoretically proposed for use in a variety of different applications \cite{marvian_generalization_2011, keyl_estimating_2001, hayashi_quantum_2002, eisert_classical_2000, bartlett_classical_2003,plesch_efficient_2010}.  
In this paper we develop a practical scheme for implementing the QSWT, and experimentally demonstrate it with photonic qubits, compressing a three-qubit ensemble into two qubits.
(The compression of a quantum ensemble is very different from, and should not be confused with, quantum source-coding \cite{schumacher_quantum_1995, mitsumori_experimental_2003}.)

A three-qubit QSWT will compress an ensemble of three qubits into two ($\log_2[3+1]$) qubits, so that one qubit can be discarded without information loss. 
A quantum circuit implementing the three-qubit QSWT is shown in figure 1a.  
In this circuit, the two single-qubit unitaries, $\hat{U}_1$ and $\hat{U}_2$, are defined so that $\hat{U}_1\left(\sqrt{\frac{2}{3}}\ket{0}+\sqrt{\frac{1}{3}}\ket{1}\right)=\ket{0}$, $\hat{U}_2\left(\sqrt{\frac{1}{3}}\ket{0}+\sqrt{\frac{2}{3}}\ket{1}\right)=\ket{0}$ and $\hat{U}_2\hat{U}_1=\hat{X}$.
It is straightforward to show that if the three input qubits are prepared in $\ket{\psi}=\alpha\ket{0}+\beta\ket{1}$ the output will be transformed into $\ket{\phi}_{1,2}\ket{0}_{3}$, where
\begin{equation}
\ket{\phi}_{1,2} = \alpha^3\ket{00}+\sqrt{3}\alpha^2\beta\ket{01}+\sqrt{3}\alpha\beta^2\ket{10}+\beta^3\ket{11}.
\end{equation}
Since the third qubit is always in $\ket{0}$ this circuit unitarily maps all of the information onto the first two qubits.  
(Such circuits can be efficiently made for any value of N, requiring one to keep only $\log_2[N+1]$ qubits \cite{bacon_efficient_2006, plesch_efficient_2010}.)  
In the case of identical pure-state qubits the final two disentangling gates (of the circuit in figure 1a) can be implemented using measurement and feed-forward, as shown in figure 1b \cite{Griffiths_semiclassical_1996}.  
Now qubit 3 is measured and an operation is performed on the first two qubits which depends on this result.  
This simplification produces $\ket{\phi}_{1,2}$, and thus performs as well as the full QSWT (see the Supplemental Material for a full derivation of this).

To understand why the compression of an ensemble of three identical qubits into two does not lose information, consider how one would estimate $\expec{\hat{Z}}$ (which we will refer to as $Z_\mathrm{true}$, the ``true value'' of this expectation value) with and without quantum data compression.
In short, without compression each qubit is measured in the same basis and an estimate is calculated from a tally of the number of spin-up and spin-down measurement results.
This tally is an integer between $0$ and $N$, and can therefore be written as a $(\log_2[N+1])$-bit string.
Explicitly, $\hat{Z}$ is measured on the three qubits, and $Z_\mathrm{true}$ is estimated directly from the individual outcomes $Z_i=\pm 1/2$ as $Z_\mathrm{direct}=(Z_1+Z_2+Z_3)/3$.
$Z_\mathrm{direct}$ has four possible values, given by the number of spin-up measurement results, which can be 3, 2, 1, or 0, corresponding to \textit{maximum-likelihood} estimates of $Z_\mathrm{direct}=\{+1/2,+1/6,-1/6,-1/2\}$, respectively.  
Since the permutation information (which specific qubits came out spin-up or spin-down) is irrelevant there are $N+1$ (four) rather than $2^N$ (eight) outcomes.  
The QSWT removes this irrelevant permutation information, compressing an $(N+1)$-valued outcome from a $(2^N)$- into an $(N+1)$-dimensional system.  
Quantum data compression amounts to encoding this information directly in $(\log_2[N+1])$ qubits, discarding the rest; so as long as the coherence between all such states is preserved, the resulting quantum state faithfully preserves the statistics of such tallies \emph{in all bases}.
With quantum data compression, $Z_\mathrm{true}$ is estimated by measuring both compressed qubits and computing $Z_\mathrm{comp}=(2Z_1+Z_2)/3$ (which can take the same four values as $Z_\mathrm{direct}$).  
To quantify the quality of the two estimates, $Z_\mathrm{direct}$ and $Z_\mathrm{comp}$, we compare their statistical variances (since the expectation values of both estimates are equal to $Z_{\mathrm{true}}$, their variances are equivalent to their mean-squared error).  
For the single-qubit state $\alpha\ket{0}+\beta\ket{1}$, both $Z_\mathrm{direct}$ and $Z_\mathrm{comp}$ have variances of ${|\alpha|^2|\beta|^2}/{3}$ (as expected, given that spin measurements obey binomial statistics).  
These identical statistics indicate that there is just as much information about $Z_\mathrm{true}$ in the two compressed qubits as there is in the three uncompressed qubits.  
More importantly, measurements on the compressed state $\ket{\phi}_{1,2}$ can be used to estimate \textit{any} single-qubit operator with the same statistical uncertainty as a direct measurement (the Supplemental Material contains a full description of how to perform other measurements).
It is in this sense that the two-qubit state $\ket{\phi}_{1,2}$ carries as much information about $\ket{\psi}$ as the three-qubit input $\ket{\psi}^{\otimes 3}$.

\begin{figure}%[p!]
\includegraphics[width=\FCW]{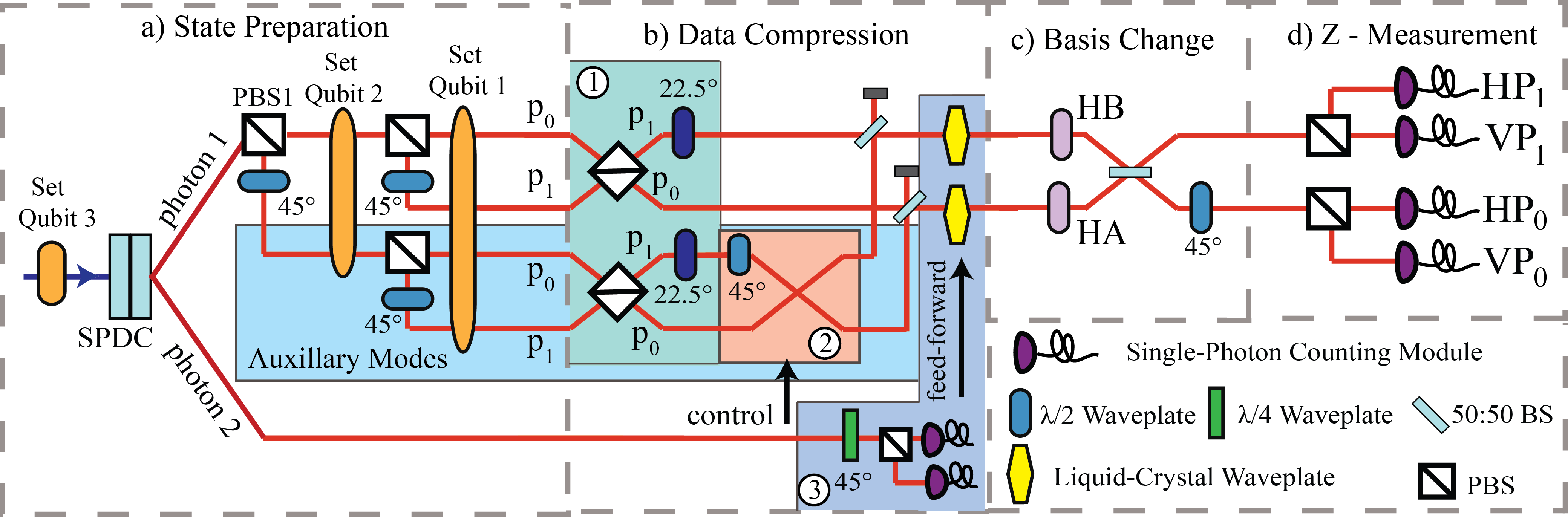}
\caption{\label{fig:2} 
{\bf Optical Implementation: ~~}
{\bf a,b) State preparation and data compression ---} Two photons, generated via spontaneous parametric down-conversion (SPDC), are used to encode three qubits. Qubit 1 is encoded in the polarization of photon 1, qubit 2 in its path degree-of-freedom (the logical paths are labelled $P_0$ and $P_1$), and qubit 3 in the polarization of photon 2 (initially entangled with an additional path degree-of-freedom of photon 1).
After the data compression circuit, only photon 1 remains, encoding a path and polarization qubit.
{\bf c,d) Measuring the compressed qubits ---}  Any single-qubit measurement can be made on the compressed state in two steps: first the basis is set (c), and then a $\hat{Z}$ measurement is performed (d). The $\hat{Z}$ measurement has four outcomes: H$P_1$, V$P_1$, H$P_0$, and V$P_1$ (where H (V) stands for horizontal (vertical) polarization).
The areas numbered 1-3 correspond to circuit elements shown in figure 1b.
}
\end{figure}

To demonstrate this protocol experimentally, we use three qubits, which we encode in the path and polarization degrees-of-freedom of two photons \cite{Fiorentino_Deterministic_2004, Fiorentino_single-photon_2005}.  
Such hybrid quantum systems, using multiple degrees-of-freedom of photons, have proven very useful for demonstrating quantum protocols \cite{barreiro_beating_2008, vitelli_quantum_2012, gao_experimental_2010}, testing fundamental issues in quantum mechanics, \cite{rozema_violation_2012, lloyd_closed_2011} and simplifying quantum logic gates \cite{lanyon_simplifying_2009,zhou_adding_2011}. 
In the circuit of figure 1b, qubit 1 is encoded in the polarization of photon 1, qubit 2 is encoded in an additional path degree-of-freedom of the same photon, and qubit 3 is encoded in the polarization of a second photon. 
After the circuit is completed, the information of all three qubits is stored in the first two logical qubits, both encoded in photon 1, allowing us to discard the second photon entirely. 
A sketch of our optical implementation is shown in figure 2, and an in-depth explanation of how it implements the quantum circuit of figure 1b is presented in the Supplemental Material. 
The two compressed qubits are encoded in the path and polarization of photon 1; to perform the post-selective disentanglement, measurements of these two qubits are post-selected on a measurement of photon 2.
This corresponds to a coincidence event between a measurement on photon 2 signalling $\ket{H+iV}/\sqrt{2}$ and any of the four detectors for photon 1.
There are four detectors because there are two possible path outcomes and two possible polarization outcomes.
These coincidence events correspond to four different estimates of $Z_\mathrm{true}$: $HP_0=\ket{00}\implies Z_\mathrm{comp}=+1/2$, $HP_1=\ket{01}\implies Z_\mathrm{comp}=+1/6$, $VP_0=\ket{10}\implies Z_\mathrm{comp}=-1/6$, or $VP_1=\ket{11}\implies Z_\mathrm{comp}=-1/2$.

\begin{figure}%[p!]
\includegraphics[width=\FCW]{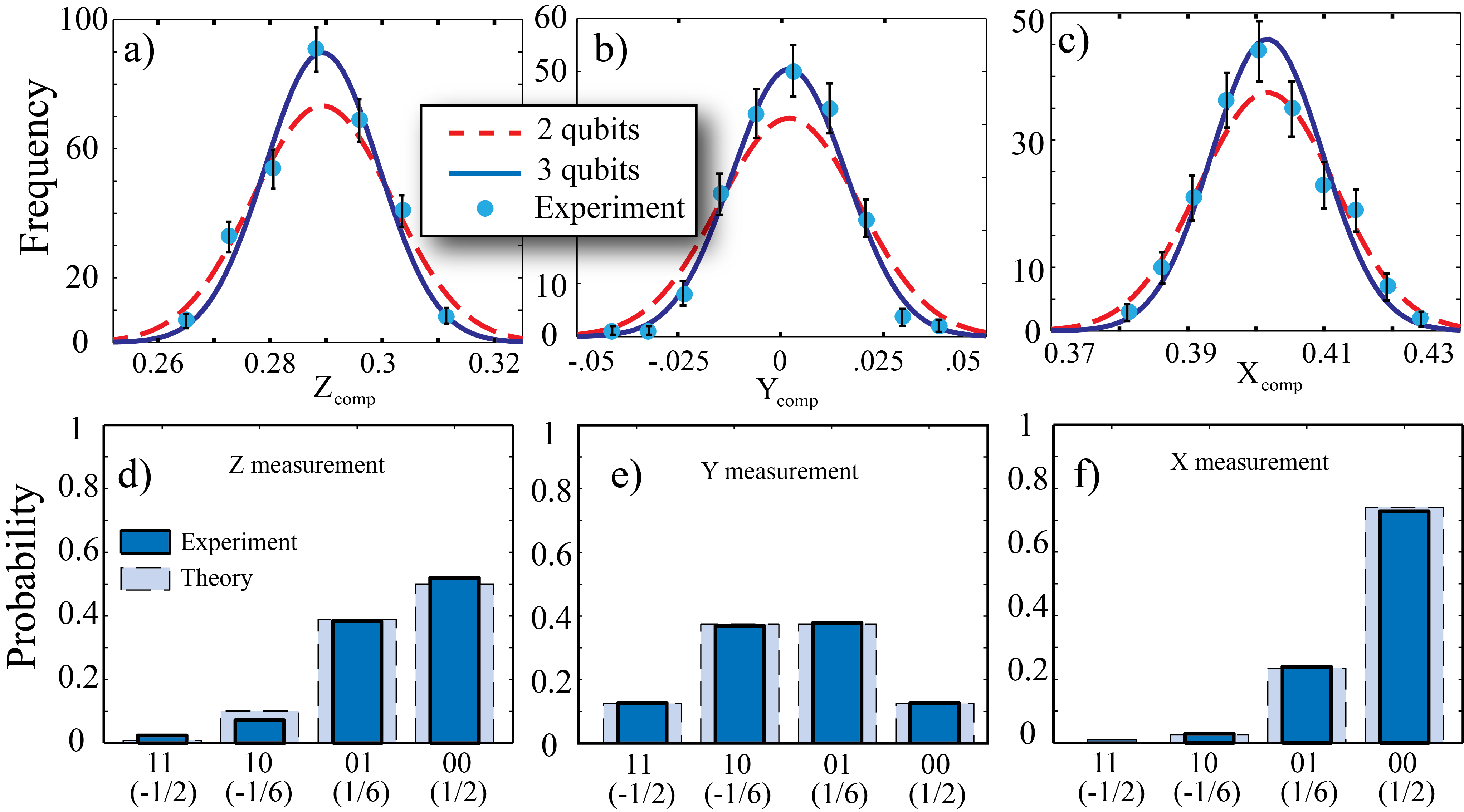}
\caption{\label{fig:3} 
{\bf Sample raw data for an input state ${\cos(2\theta)\ket{0}+ \sin(2\theta)\ket{1}}$, for $\theta=13.5^\circ$  ---} {a)-c)} Histograms of estimates the spin along $\hat{Z}$, $\hat{Y}$ and $\hat{X}$, after $M$ trials (defined in the text) of the data compression circuit.  
The bars are experimentally measured data, and the blue (red) curve is a normal distribution of width $V_1/3M$ ($V_1/2M$) normalized to have the same area as the experimental histogram, where $V_1$ is the single-qubit variance.
{Error bars are calculated using a Monte Carlo simulation of the measurement scheme that is described in the text.}
{d)-f)}  The experimentally observed probabilities for measuring the two qubits and finding them in $\ket{00}$, $\ket{01}$, $\ket{10}$, or $\ket{11}$ for $\hat{Z}$, $\hat{Y}$ and $\hat{X}$ measurements.
The dark blue bars are the experimentally measured counts, normalized by the total number of counts, and the light bars are the theoretically predicted results.
{The error bars here are suppressed because they are not visible on the scale of the plots.}
}
\end{figure}

To test the performance of our circuit, the compressed system was measured and a number of representative single-qubit observables were estimated.  
For each measurement, the two qubits were found in one of four states, corresponding to expectation-value estimates of +1/2, +1/6, -1/6, or -1/2.
Since a single measurement does not yield information about the statistical performance of our circuit, we ran the circuit many times for the same input state and final measurement.  
The number of runs was typically $M\approx500$.  
For each run, $\hat{S}$ (either $\hat{X}$, $\hat{Y}$, or $\hat{Z}$) was measured on the output and the spin expectation value was estimated as ${S}_\mathrm{comp}=(2S_1+S_2)/3$, then the average of ${S}_\mathrm{comp}$ over runs was calculated.  
This entire process formed a single trial, and was repeated about 250 times. 
The resulting distributions of the averages of ${S}_\mathrm{comp}$ are plotted in figure 3a-c for $\hat{S}$ = $\hat{X}$, $\hat{Y}$, and $\hat{Z}$ with the initial single-qubit state $\cos(2\theta)\ket{0}+ \sin(2\theta)\ket{1}$ and $\theta=13.5^\circ$. 
If each of the $M$ measurements encodes the information of three qubits (as we expect) the distribution should have a variance given by the single-qubit variance ($V_1=\cos^2(2\theta)\sin^2(2\theta)$) divided by the total number of qubits sampled: $3M$, three times the number of runs in each trial.
This prediction is shown in blue on figures 3 a-c. 
On the other hand, a measurement made on two independent qubits would exhibit a variance of $V_1/(2M)$, $1.5$ times larger; this distribution is shown in red for comparison. 
The narrower blue curve, describing the behaviour of three qubits, is a much better fit to our observed data than the red curve, indicating that the amount of information extractable from the two compressed qubits is close to the full information present in the three original qubits.

\begin{figure}%[p!]
\includegraphics[width=\FCW]{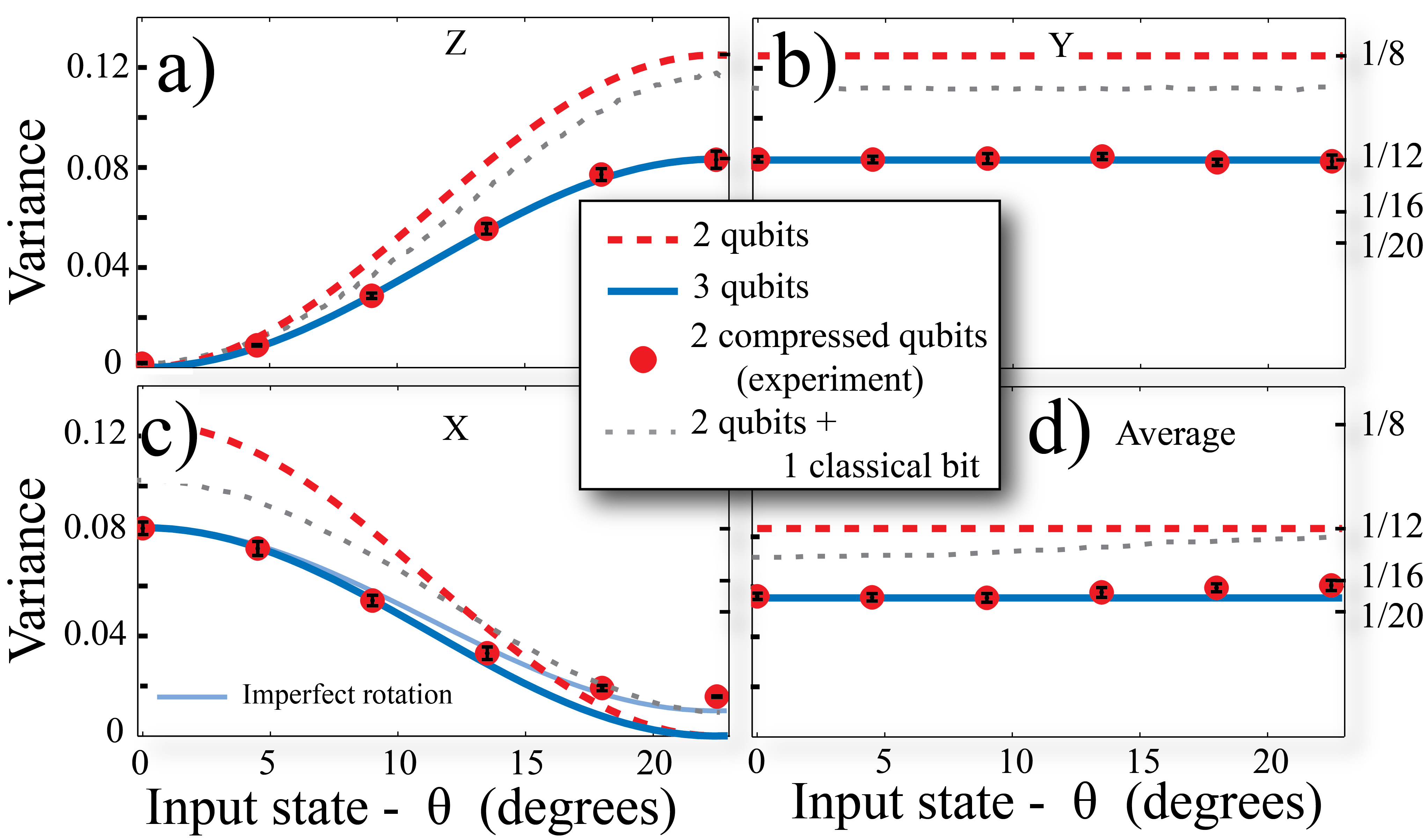}
\caption{\label{fig:4}
{\bf Measurement Variances for various input states ---}  The solid blue lines are the theoretical variances resulting from performing a measurement on three independent qubits (and thus our two compressed qubits), the dashed red lines are for two independent qubits, the grey dashed lines are the theoretical variance when two independent qubits are measured optimally and a random measurement is performed on a third qubit, and the circles are the variances which we observe when experimentally measuring the two compressed qubits. 
a-c) By sending in various different input states, parametrized by $\theta$ as $\cos 2\theta\ket{0}+\sin 2\theta\ket{1}$, we see that the two compressed qubits demonstrate the statistics of three independent qubits for $\hat{Z}$, $\hat{Y}$ and $\hat{X}$ measurements. 
d) Averaging the variances of $\hat{Z}$, $\hat{Y}$ and $\hat{X}$ yields the variance averaged over all possible measurements.
}
\end{figure}

To further quantify the performance of our compression circuit, we measure the `single-shot' distributions of ${X}_\mathrm{comp}$, ${Y}_\mathrm{comp}$, and ${Z}_\mathrm{comp}$. 
To do this we again prepare each of the three input qubits in $\cos(2\theta)\ket{0} + \sin(2\theta)\ket{1}$, run our circuit, measure one of the observables $\hat{X}$, $\hat{Y}$, or $\hat{Z}$ (each measurement results in an estimate of +1/2, +1/6, -1/6, or -1/2) and bin the results. 
For each observable the circuit was run approximately $10^5$ times. 
The resulting normalized distributions are plotted in figure 3d-f for $\theta=13.5^\circ$. 
We observe very good agreement between our experimental data (dark bars) and theory (larger light bars).  
Next, we vary the input states, preparing a range of $\theta$ values, and measure the variance of the resulting single-shot distributions of ${X}_\mathrm{comp}$, ${Y}_\mathrm{comp}$, and ${Z}_\mathrm{comp}$ for each input state.
These experimentally-measured variances are the circles, plotted versus $\theta$, in figure 4a-c.
The curves in figure 4a-c are theory corresponding to the variance of two independent qubits $V_1/2$ (red dashed curve) and three independent qubits $V_1/3$ (blue solid curve).
For all but two data points in the $\hat{X}$ measurement (discussed in the Supplemental Material), our experimental data agree very well with the three-qubit variance.
In addition to these three observables, one would ideally quantify the variance averaged over all possible measurements.
Such a measurement would indicate how much information could be extracted about arbitrary measurements.
Conveniently, for a given state, this average measurement variance is the same as simply averaging the variances of $\hat{X}$, $\hat{Y}$ and $\hat{Z}$.
(That is to say that the uniformly distributed discrete subensemble $\{ \hat{X},\hat{Y},\hat{Z} \}$ is an \emph{averaging set} for the SO(3) uniformly (Haar) distributed superensemble $\{ \hat{S}(\theta,\phi) \}$ for variance \cite{seymour_averaging_1984}; we derive this in the Supplemental Material).  
The resulting averaged variance, for a given state, is plotted in figure 4d.  
This clearly demonstrates that our circuit compresses three qubits into two, and we can conclude that all of the compressed states we tested faithfully encode the information about any single-qubit measurement.

So far we have imagined that, if presented with three qubits and a two-qubit quantum memory, our strategy in the absence of a compression circuit would be to store two of the qubits and discard the third.  
This measurement scheme has a variance $1.5$ times larger than we obtain with compression (red curve in figure 4). 
A better approach would be to measure the third qubit before discarding it.  
The classical bit obtained would provide extra information and could be combined with the subsequent measurement of the two stored qubits in the correct basis, yielding an improved estimate of the single-qubit spin.  
Any compression algorithm should be compared to such a strategy, in order to quantify the performance given a limited amount of quantum memory, without placing unreasonable constraints on the classical memory.
We analyze this protocol in the Supplemental Material; the result is the dotted grey curve in figure 4.  
Our compression scheme outperforms even this improved protocol.

Finally, it is worth mentioning that our techniques could be useful beyond compressing sets of identical input states.  
For instance, one could also exponentially compress any permutationally invariant pure state.  
Permutationally invariant states include several entangled states which have been shown to be invaluable for quantum communication and quantum computing \cite{adamson_multiparticle_2007, adamson_detecting_2008, toth_permutationally_2010}, including GHZ states\cite{greenberger_going_1989,bouwmeester_observation_1999} and W-states\cite{durr_Three_2000}.
%, and cluster states\cite{Raussendorf_one-way_2001,walther_Experimental_2007}.  
Many other applications of the QSWT, outside of compression, exist \cite{marvian_generalization_2011, keyl_estimating_2001, hayashi_quantum_2002, eisert_classical_2000, bartlett_classical_2003, plesch_efficient_2010}, and for some applications our feed-forward simplification performs optimally. 
Given the exponential reduction in the size of the required quantum memory, and the many applications of the QSWT, circuits such as the one we have demonstrated hold great promise for both future quantum computing and quantum communication architectures.

\begin{acknowledgments}
We thank Robin Blume-Kohout for helpful discussions, and Alan Stummer for help building our coincidence circuit.  LAR, DHM, AH, and AMS acknowledge support from the Natural Sciences and Engineering Research Council of Canada and the Canadian Institute for Advanced Research, and PST acknowledges the support of a \textit{wakatehake} travel grant from the University of Tokyo.

\noindent{$^{*}$ Corresponding author, lrozema@physics.utoronto.ca.}

\noindent{$^{\dagger}$ Now at the School of Physics, H. H. Wills Laboratory, University of Bristol, Bristol, UK BS81TL}

\end{acknowledgments}

\bibliography{experimental_QDC_draft13.0}
\bibliographystyle{Science}

\newpage

\section*{\bf Supplemental Material} 

\section{Additional Experimental Details} 

\subsection{State Preparation} In our implementation, qubit 1 is encoded in the polarization of photon 1, qubit 2 in the path of the same photon, and qubit 3 in the polarization of a second photon (figure 2a of the main text).  
We generate photon pairs using a type-I spontaneous parametric down-conversion (SPDC) source in a ``sandwich-configuration'' \cite{kwiat_ultrabright_1999} (each crystal is 1mm of BBO, and they are pumped by 500mw of 404nm light, generated by frequency-doubling 808nm light from a femtosecond Ti:Sapph laser, using a 2mm-long BBO crystal). 
Our source creates photons in the entangled state $\alpha\ket{HH}_{1,2}+\beta\ket{VV}_{1,2}$ with a fidelity of $\approx94\%$, measured with standard two-photon polarization tomography.  The amplitudes $\alpha$ and $\beta$ are controlled via the pump polarization.

This polarization entanglement is converted into entanglement between the polarization of photon 2 and an ``auxiliary'' path degree of freedom of photon 1, by passing photon 1 through a polarizing beamsplitter (PBS) followed by a half-waveplate (HWP) at $45^\circ$ in the reflected port.  
After this, the state of the system is $\ket{H}_1(\alpha\ket{a_0}_{1}\ket{H}_2+\beta\ket{a_1}_{1}\ket{V}_2)$, where $\ket{a_1}$ ($\ket{a_0}$) refers to photon 1 being in the auxiliary path (or not).
Qubit 3 is the polarization state of this second photon (whose state, defined by $\alpha$ and $\beta$, is set by the pump polarization), and it is entangled with the auxiliary mode of photon 1.
This entanglement is later used to implement quantum-logic gates between qubits encoded in photon 1 and photon 2.
Since photon 1 is now horizontally polarized, the state of qubit 2 (the path of photon 1) can be set by setting the polarization of photon 1 and converting it to a path qubit with a PBS and a HWP.  
Finally, the state of qubit 1 (the polarization of photon 1) can be set.
This entire procedure results in the state 
\begin{eqnarray}
(\alpha\ket{H}_1+\beta\ket{V}_1)\otimes(\alpha\ket{p_0}_{1}+\beta\ket{p_1}_{1})\\ \nonumber
\otimes(\alpha\ket{a_0}_{1}\ket{H}_2+\beta\ket{a_1}_{1}\ket{V}_2),
\end{eqnarray}
where $\ket{p_0}_{1}$ and $\ket{p_1}_{1}$ refer to the state of the path qubit encoded in photon 1.

\subsection{Logic gates} The quantum circuit that we implement (figure 1b of the main text) can be broken into three parts: the two-qubit QSWT (box 1), the controlled gates between qubit three and the first two qubits (box 2), and the post-selective disentangling operation (box 3).
With our encoding, the gates labelled ``two-qubit QSWT'' can be performed deterministically using linear optics.  
The controlled-not (CNOT), with the polarization qubit as the control and the path qubit as the target, is implemented by using a PBS to swap the path modes only when the photon is vertically polarized (which we define as $\ket{1}$ for the polarization qubits); the controlled-Hadamard, with path as control and polarization as target, is implemented by using a half-waveplate at $22.5^\circ$ to rotate the polarization only if the photon is in path 1 (defined as the $\ket{1}$ state of the path qubit).
These optical elements are shown in shaded area 1 of figure 2 in the main text.

The next two gates (box 2) are experimentally more challenging, requiring photon 2's polarization to modify the path and polarization of photon 1. 
As before, the ``uncontrolled'' implementations of these gates can be constructed between the path and polarization qubits encoded in photon 1 by using linear optics:  
the NOT gate on the path qubit is implemented by swapping the path modes, and the CNOT gate, with path qubit as the control and the polarization qubit as the target, is achieved by placing a HWP at $45^\circ$ only in path 1.
To be clear, the CNOT gate that we are referring to at this point is an ``uncontrolled implementation'' of the three-qubit Toffoli gate.
The challenge comes in conditioning them on the polarization state of photon 2 (qubit 3).  
Zhou \textit{et al.} showed that this can be conveniently achieved by using ``controlled-path'' gates\cite{zhou_adding_2011} to ``take a shortcut through a higher dimension'' \cite{lanyon_simplifying_2009}.

A controlled-path gate places qubit 2 in an auxiliary mode dependent on the state of qubit 1.  
It is essential that the controlled-path gate place qubit 2 in the auxiliary mode coherently, so that if qubit 1 is in a superposition of $\ket{0}$ and $\ket{1}$ entanglement will be generated between the the state of qubit 1 and the mode of qubit 2.
If this is the case, then placing some gate $\hat{A}$ in the auxiliary mode and recombining the auxiliary mode with the original mode using another controlled-path gate will create entanglement between the state of qubit 1 and $\hat{A}$ either being applied to qubit 2 or not.
In other words, this process has implemented a controlled-$\hat{A}$ gate with qubit 1 as the control and qubit 2 as the target (see figure 5a of the Supplemental Material).

\begin{figure}
\includegraphics[scale=.5]{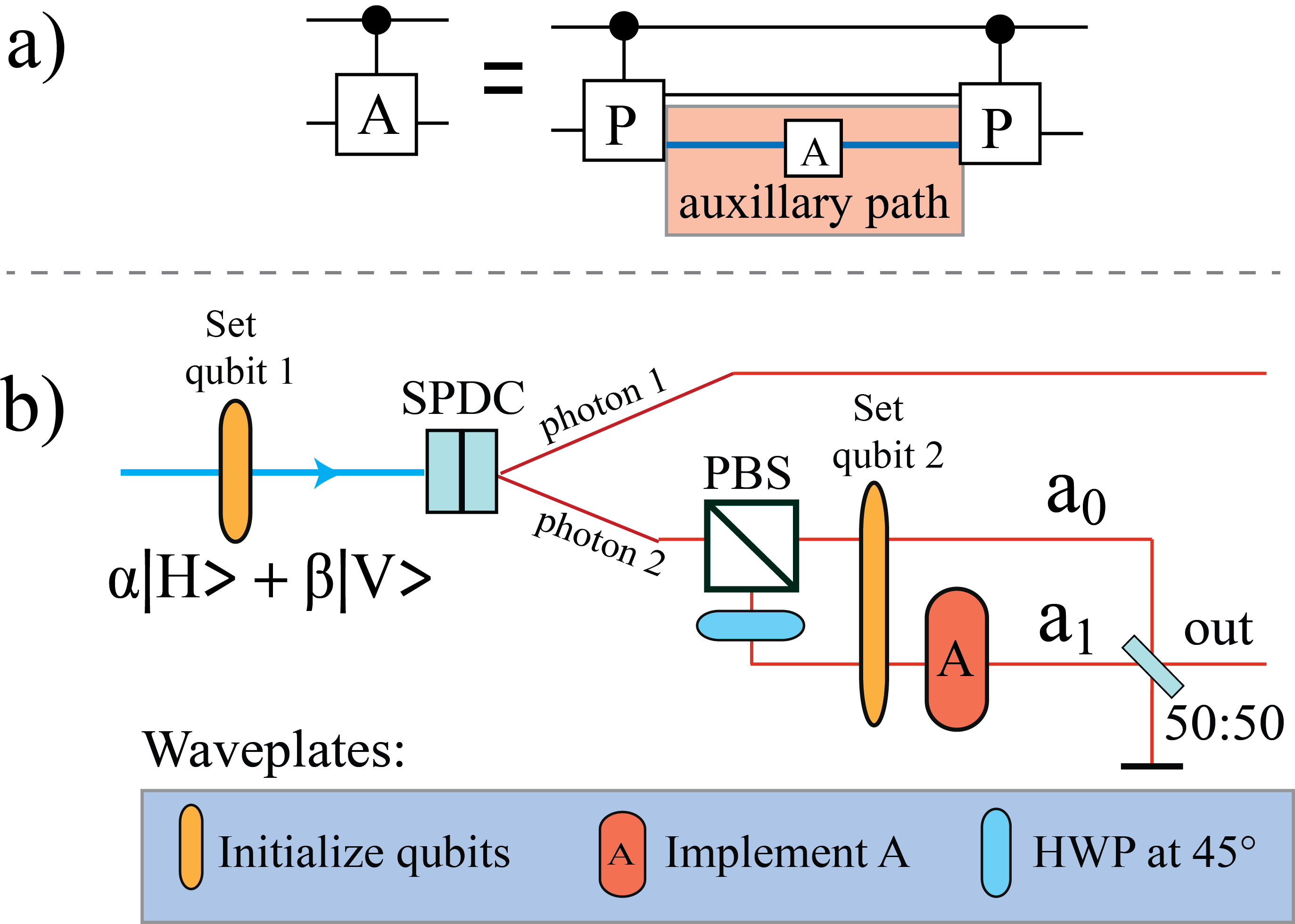}
\caption{\label{fig:6} 
\textbf{ a) Controlling Arbitrary Gates ---} Controlled-path gates, gates which move the target qubits to auxiliary modes, can be used to add control to any quantum gate by placing the gate only in the auxiliary mode.
{\bf b) Entanglement-Driven Controlled-Unitary for Polarization Qubits ---} 
The controlled path gate can be replaced by existing entanglement between the control qubit and the target qubit.  
Entanglement is generated between the polarization of two photons, which is then converted into entanglement between the upper photon's polarization and the lower photon's path.  
The input state of the upper photon is set by setting the pump polarization, and the state of the second photon is set with waveplates after the polarizing beamsplitter. 
The gate we wish to control is placed only in the auxiliary mode, and the modes are recombined using a 50:50 beamsplitter.}
\end{figure}

Since controlled-path gates could be just as challenging to implement as a CNOT gate, Zhou \textit{et al.} went on to show that this scheme could be simplified by using prior entanglement between the first qubit and the auxiliary mode of the second qubit.
Replacing a controlled-path gate with prior entanglement is possible because the crucial effect of the controlled-path gate is the entanglement that it generates between the state of qubit 1 and the mode of qubit 2.
This is an extremely useful technique for photonic logic gates since it is often easier to generate entanglement from a photon source than it is to generate entanglement between independent photons.

To understand how this works in practice, consider implementing an entanglement-driven controlled-$\hat{A}$ gate between two polarization qubits (figure 5b of the Supplemental Material).
The polarization of photon 1 will be the control qubit and the polarization of photon 2 will be the target qubit.
To do this, we start by generating the polarization-entangled photons of the form $\alpha\ket{HH}_{1,2}+\beta\ket{VV}_{1,2}$ from a down-conversion source; $\alpha$ and $\beta$ are set by setting the pump polarization.
$\alpha$ and $\beta$ will be used to define the state of the first qubit: $\ket{\psi_1}=\alpha\ket{H}_1+\beta\ket{V}_1$.
Next, we use a PBS to create an auxiliary path for photon 2, and use a half waveplate at $45^\circ$ to reset the polarization of photon 2 to $\ket{H}$.
At this point the state is $(\alpha\ket{H}_{1}\ket{a_0}_2+\beta\ket{V}_{1}\ket{a_1}_2)\ket{H}_{2}$, where $\ket{a_0}$ and $\ket{a_1}$ refer to the mode of photon 2.
Then the polarization state of the second photon is set to $\ket{\psi_2}=\gamma\ket{H}_2+\delta\ket{V}_2$ using waveplates common to both modes, which allows any separable input state of the two photons' polarization to be prepared.
This completes the state preparation with:
$(\alpha\ket{H}_{1}\ket{a_0}_2+\beta\ket{V}_{1}\ket{a_1}_2)\ket{\psi_2}_{2}$.
Note that the first qubit $\ket{\psi_1}$ is already entangled with the path mode of the second qubit, as if the the CP gate has already been applied.
Now if polarization optics (implementing $\hat{A}$) are placed only in the mode $a_1$ they will be controlled in effect by the polarization of photon 1: $\alpha\ket{H}_{1}\ket{a_0}_2(\ket{\psi_2}_{2})+\beta\ket{V}_{1}\ket{a_1}_2(\hat{A}\ket{\psi_2}_{2})$.
Finally the auxiliary modes are post-selectively recombined at a 50:50 beamsplitter, such that if photon 2 is found at the output, then the state of the two photons is $\alpha\ket{H}_{1}(\ket{\psi_2}_{2})+\beta\ket{V}_{1}(\hat{A}\ket{\psi_2}_{2})$.
Thus we are able to implement a controlled-$\hat{A}$ gate using entanglement created from the source, while being able to prepare any separable input state of the the two qubits.

In our experiment, since the polarization of our photons is entangled, we can use PBS1 in figure 2a (of the main text) to create an auxiliary set of paths for photon 1, so optics placed only in these modes (shaded area 2 in figure 2) are effectively controlled by the polarization of photon 2.

The final step of our data compression circuit is measurement and feed-forward to disentangle qubit 3 (illustrated in shaded area 3).  
In Section III, we show that this is possible by measuring qubit 3 and applying a unitary on the first two qubits based on the result.  
Experimentally, this is accomplished by measuring the polarization of photon 2 in the circular basis, and applying birefringent phases on photon 1 with liquid-crystal waveplates (LCWPs) set to $0^\circ$.  
Ideally, the LCWP retardances are switched dependent on the measurement outcome of photon 2.  
Since our LCWPs are not fast enough we set them to correct the phases only when photon 2 is projected onto $(\ket{H}+i\ket{V})/\sqrt{2}$ and discard the other case. 
With faster feed-forward (using Pockels cells, for example), both cases could be corrected \cite{prevedel_high-speed_2007}.

After the compression, we are left with a single photon encoding the two compressed qubits.  
The path and polarization of this photon are measured (figure 2c-d of the main text) in coincidence with the polarization measurement on photon 2.  
After post-selection of photon 2 in the state $\ket{H+iV}/\sqrt{2}$, approximately 1000 events per second are observed.

\subsection{Performance} The net result of our implementation is a series of four nested interferometers when measuring $\hat{Z}$, and five when measuring $\hat{X}$ or $\hat{Y}$.  
The phase of each interferometer was measured to be stable for at least five minutes (drifting less than $1\%$) so that, with our detection rates, we could collect sufficient data in one minute without significant phase drift occurring.  
The first four interferometers had visibilities $>98.5\%$, while that of the fifth was $97.4\%$.  
This can be interpreted as an error in the measurement basis setting.  
This is because rather than measuring $\ket{+x}\bra{+x}$, we measured $(1-p)\ket{+x}\bra{+x} + p\ket{-x}\bra{-x}$, where $p$ is the leakage into the interferometer's dark port.  
We measured $p=0.015$ and used this to simulate the effect on the variance (thin blue curve in figure 3c of the main text), which describes our experimental data well.
This led to the deviation of variance of $\hat{X}$ as $\theta$ approached $22.5^\circ$.
The $\hat{X}$-measurement was the most sensitive to this error because to measure a variance of zero (as predicted by theory), all of the photons had to have exited the final bright port, and even a small amount of leakage into the ``dark-port''  would increase the variance.
For the states that we used, the $\hat{Y}$-measurement should never have resulted in a dark port, and was thus not sensitive to this error, while the $\hat{Z}$-measurement was made without this final interferometer in place.

\section{Measuring the compressed system } 
After the initial three-qubit state is compressed into two qubits the information needs to be read out. 
Since the compression is unitary, one could run an inverse QSWT with an ancillary qubit initialized in $\ket{0}$. 
This would recreate the initial three qubit state. 
This is unnecessary, aside from being experimentally challenging.
We instead show that the desired observables may be measured directly on the compressed qubits. 
As discussed in the text, measuring $\hat{Z}$ on each of the two compressed qubits and interpreting the result as a 2-bit number yields the same information as measuring $\hat{Z}$ on the input three qubits, the 2-bit number being equivalent to a tally of how many of the three qubits were found in $\ket{0}$.
As we will see below, other spin measurements are necessarily non-local, but still feasible to implement on the compressed qubits.

It is possible to map the basis states of the output qubit pair onto the four basis states of an effective spin-3/2 system formed by the symmetric states of the three-qubit input.
This results naturally from adding the angular momentum of the three input qubits (spin-1/2 particles). 
To do this we relabel the basis states as: $\ket{00}\rightarrow\ket{m=+3/2}$, $\ket{01}\rightarrow\ket{m=+1/2}$, $\ket{10}\rightarrow\ket{m=-1/2}$, and $\ket{11}\rightarrow\ket{m=-3/2}$. 
In order to estimate $\expec{\hat{Z}}$, we measure the $m$ value of the effective spin-3/2 system (by measuring $\hat{Z}$ on the two qubits) and report $m/3$ as our estimate.  
Explicitly, this means that finding the two compressed qubits in $\ket{00}=\ket{m=+3/2}$ corresponds to finding the effective particle to be spin-up along $\hat{Z}$, resulting in an estimate of $+1/2$ for the expectation value.  
Similarly, finding $\ket{01}=\ket{m=+1/2}$ would yield an estimate of $+1/6$, and so on.  
It can now be seen why measuring, say, $\hat{X}$ \textit{locally} on each of the compressed qubits does not yield information about $\expec{\hat{X}}$: we must measure $\hat{X}$ collectively on the spin-3/2 particle, not on the individual compressed qubits. 
Just as with any spin system, this can be accomplished by changing the measurement basis of the effective spin-3/2 particle, and then measuring $\hat{Z}$ locally.

The basis change can be calculated for any measurement but its implementation is non-trivial, requiring entangling gates between the two compressed qubits, in general.  
For example, measuring $\hat{X}$ requires the (entangling) transformation shown in equation \ref{eq:basis} to be applied to the compressed qubits before locally measuring $\hat{Z}$ on each qubit.
\begin{eqnarray}
\label{eq:basis}
2\sqrt{2}\ket{00}\rightarrow\left(\ket{0}+\sqrt{3}\ket{1}\right)\ket{0} + \left(\sqrt{3}\ket{0}+\ket{1}\right)\ket{1}\nonumber\\
2\sqrt{2}\ket{01}\rightarrow\left(\sqrt{3}\ket{0}-\ket{1}\right)\ket{0} + \left(\ket{0}-\sqrt{3}\ket{1}\right)\ket{1}\nonumber\\
2\sqrt{2}\ket{10}\rightarrow\left(\sqrt{3}\ket{0}-\ket{1}\right)\ket{0} - \left(\ket{0}-\sqrt{3}\ket{1}\right)\ket{1}\nonumber\\
2\sqrt{2}\ket{00}\rightarrow\left(\ket{0}+\sqrt{3}\ket{1}\right)\ket{0} - \left(\sqrt{3}\ket{0}+\ket{1}\right)\ket{1}.
\end{eqnarray}
In our experiment, the two compressed qubits are stored in the path and polarization degrees of freedom of a single photon, so we can deterministically implement the entangling gates required for the basis change using only linear optics, in a manner similar to the implementation of the path/polarization logic gates discussed in the Methods section.  
We implement this by placing different waveplates in the two paths of the photon, and combining the two paths at a 50:50 beamsplitter. 
The basis change required for a $\hat{Y}$ measurement can be implemented by simply changing the waveplate settings.

\section{Measurement and Feed-forward} 

The full quantum circuit of figure 1a of the main text is not necessary if the input qubits are guaranteed to be identically prepared. 
In this case the final two gates can be replaced by measurement and feed-forward (figure 1b of the text). 
To understand this, consider the state before the disentangling gates (for input qubits in state $\alpha\ket{0}+\beta\ket{1}$):
\begin{eqnarray}
\alpha^3\ket{00}\ket{0}+\sqrt{3}\alpha^2\beta\ket{01}\ket{\phi_1}+\sqrt{3}\alpha\beta^2\ket{10}\ket{\phi_2}\\ \nonumber +\beta^3\ket{11}\ket{1},
\end{eqnarray}
where $\ket{\phi_1}=\sqrt{\frac{2}{3}}\ket{0}+\sqrt{\frac{1}{3}}\ket{1}$ and $\ket{\phi_2}=\sqrt{\frac{1}{3}}\ket{0}+\sqrt{\frac{2}{3}}\ket{1}$. 
If at this point qubit 3 is simply discarded (traced over) the first two qubits will be left in a mixed state, and information is lost. 
If instead of being discarded, qubit 3 is measured in the basis $(\ket{0}\pm i\ket{1})/ \sqrt{2}$, the first two qubits are left in one of two possible states with equal probability, dependent on this outcome. 
If qubit 3 is found in $(\ket{0}+ i\ket{1})/ \sqrt{2}$ the first two qubits will collapse into the state
\begin{equation}
\label{eq:collapse1}
\alpha^3\ket{00}+e^{-ia}\sqrt{3}\alpha^2\beta\ket{01} - e^{ia}\sqrt{3}\alpha\beta^2\ket{10}+\beta^3\ket{11},
\end{equation}
whereas if qubit 3 is found in $(\ket{0}- i\ket{1})/ \sqrt{2}$ the first two qubits will be left in
\begin{equation}
\label{eq:collapse2}
\alpha^3\ket{00}+e^{ia}\sqrt{3}\alpha^2\beta\ket{01} - e^{-ia}\sqrt{3}\alpha\beta^2\ket{10}-i\beta^3\ket{11},
\end{equation}
where $\exp(i a)\sqrt{3} ={\sqrt{2} + i}$.  
As can be seen in equations \ref{eq:collapse1} and \ref{eq:collapse2}, the state of the first two qubits still encodes the population information correctly, but there are extra phases which differ depending on the result of the  measurement performed on qubit 3.  
Importantly, these extra phases are independent of the input state ($\alpha$ and $\beta$).  
This allows the phases to be corrected based on the measurement outcome, without knowing in advance the input of the circuit.  
Since this will prepare the ideal state (equation 1 of the main text) on the first two qubits, this simplified measurement-and-feed-forward scheme must perform just as well as the full QSWT.

\section{Maximum-Likelihood Estimation without Data Compression}

In this section we will consider the scenario wherein one initially has three qubits, no compression circuit, and a two-qubit quantum memory. 
In this case, rather than discarding the third qubit, one could measure it in some basis, and the classical outcome stored along with the other two qubits, which would be measured optimally later when measurement axis is known. 
It is easiest to explain our analysis of this scenario in terms of a game between Alice and Bob.

Imagine that Alice prepares three qubits and gives them to Bob. 
The state that Alice prepares is unknown to Bob and known by Alice. 
Sometime later, Alice is going to ask Bob to predict the value of a spin measurement along a random direction. 
If Bob can only store two qubits, his best option is to perform our data compression algorithm and store all of the quantum information in memory. 
If he is not able to perform data compression, he could still gain some information by measuring one qubit before he discards it. 
Given this additional classical bit of information (the outcome of his spin measurement), he must come up with an estimate of the spin about some other axis that Alice is going to tell him. 
The procedure we imagine Bob following is to measure the first qubit randomly (since he has no directional information with which to make his choice), then when Alice tells him what axis she is interested in, Bob will measure the remaining two qubits along her axis. 
From these three measurement results and the knowledge of his single qubit measurement direction, he will construct and maximize a likelihood function.

To compare to our experiment, imagine that Alice prepares three qubits in $\ket{\psi}=\cos{2\theta}\ket{0}+\sin{2\theta}\ket{1}$, and then asks Bob to estimate the spin along $\hat{Z}$.
(We do not {lose generality by considering only states with a relative phase of zero because} we can simply consider different measurements; i.e. preparing states with zero phase and measuring $\hat{X}$ will behave the same as preparing states with a phase of $\pi/2$ and measuring $\hat{Y}$.)
We use the convention that the eigenvalue associated with $\ket{0}$ is $+^1/_2$, and that with $\ket{1}$ is $-^1/_2$.
In the state $\ket{\psi}$, the expectation value of $\hat{Z}$ is $\expec{\hat{Z}}=\frac{1}{2}\cos4\theta$, which we will refer to as $\expec{\hat{Z}}$'s `true value', $Z_\mathrm{true}$.  
Bob's goal is to estimate $Z_\mathrm{true}$.  
For clarity, we rewrite $\ket{\psi}$ in terms of $Z_\mathrm{true}$ as $\ket{\psi}=\sqrt{^1/_2+Z_\mathrm{true}}\ket{0}+\sqrt{^1/_2-Z_\mathrm{true}}\ket{1}$.  
Bob's protocol is now to measure the first qubit's spin along a Haar-randomly chosen axis, parametrized as ${2}\hat{S}_{\delta,\epsilon} = \cos{\delta}\sin{\epsilon}\hat{X} + \sin{\delta}\sin{\epsilon}\hat{Y} + \cos{\epsilon}\hat{Z}$. 
$\hat{S}_{\delta,\epsilon}$ has spin-up and spin-down eigenstates $\ket{S_{\delta,\epsilon}=0}=\cos{\frac{\delta}{2}}\ket{0}+{ e^{i\epsilon}}\sin{\frac{\delta}{2}}\ket{1}$ and $\ket{S_{\delta,\epsilon}=1}=\sin{\frac{\delta}{2}}\ket{0}-{e^{i\epsilon}}\cos{\frac{\delta}{2}}\ket{1}$, respectively (again with eigenvalues of $\pm^1/_2$).  
One can show that, given a state with $Z_\mathrm{true}$, Bob will find his qubit to be spin-up along $\hat{S}_{\delta,\epsilon}$  (i.e. $\ket{\psi}$ will be projected  onto $\ket{{\hat S}_{\delta,\epsilon}=0}$) with probability:
\begin{eqnarray}
\label{eq:out0}
P({ \hat S}_{\delta,\epsilon}=0|Z_\mathrm{true}) = |\langle\psi\ket{{\hat S}_{\delta,\epsilon}=0}|^2 = \\ \nonumber
\frac{1}{2}+Z_\mathrm{true}\cos\delta+\frac{1}{2}\sqrt{1-4Z_\mathrm{true}^2}\sin\delta\cos\epsilon,
\end{eqnarray} 
and similarly he will find it in $\ket{{ \hat S}_{\delta,\epsilon}=1}$ with probability
\begin{eqnarray}
\label{eq:out1}
P({ \hat S}_{\delta,\epsilon}=1|Z_\mathrm{true}) =&  \\ \nonumber
\frac{1}{2}-Z_\mathrm{true}\cos\delta-&\frac{1}{2}\sqrt{1-4Z_\mathrm{true}^2}\sin\delta\cos\epsilon.
\end{eqnarray}
These two equations would typically be interpreted as the probability of getting a spin-up (${ \hat S}_{\delta,\epsilon}=0$) or a spin-down (${ \hat S}_{\delta,\epsilon}=1$) outcome given a state with a specific value of $Z_\mathrm{true}$.  
However, since Bob's task is to conclude things about $Z_\mathrm{true}$ given an experimental outcome, here we will view equations \ref{eq:out0} and \ref{eq:out1} as the likelihood that the state had $Z_\mathrm{true}$, given that a specific outcome (${ \hat S}_{\delta,\epsilon}=0$ or ${ \hat S}_{\delta,\epsilon}=1$) was observed.  
In other words, they are ``likelihood functions'' for $Z_\mathrm{true}$ given an experimental outcome: $L(Z_\mathrm{true}|{ \hat S}_{\delta,\epsilon}=0) = P({ \hat S}_{\delta,\epsilon}=0|Z_\mathrm{true})$ and $L(Z_\mathrm{true}|{ \hat S}_{\delta,\epsilon}=1) = P({ \hat S}_{\delta,\epsilon}=1|Z_\mathrm{true})$.

After Bob randomly measures one qubit, Alice will tell him to estimate the spin along $\hat{Z}$. 
Thus his next step is to measure $\hat{Z}$ on the remaining two qubits.  
Doing this will yield three possible results: he will either find one qubit spin-down (in $\ket{0}$) and one spin-up (in $\ket{1}$), both spin-up (in $\ket{0}$) or both spin-down (in $\ket{1}$). 
These outcomes will occur with probabilities:
\begin{eqnarray}
P(\hat{Z}=0,0|Z_\mathrm{true}) &=&  \cos^4 2\theta  = \left(\frac{1}{2} + Z_\mathrm{true}\right)^2,\\
P(\hat{Z}=0,1|Z_\mathrm{true}) &=&  \, 2\cos^2 2\theta\sin^2 2\theta \,  \\ \nonumber
&=& \frac{1}{2} - 2Z_\mathrm{true}^2,\\
P(\hat{Z}=1,1|Z_\mathrm{true}) &=&  \sin^4 2\theta  = \left(\frac{1}{2} - Z_\mathrm{true}\right)^2,
\end{eqnarray}
respectively.  
He can construct likelihood functions for $Z_\mathrm{true}$, given each of these outcomes as before: $L(Z_\mathrm{true}|\hat{Z}=0,0) = P(\hat{Z}=0,0|Z_\mathrm{true})$, $L(Z_\mathrm{true}|\hat{Z}=0,1) = P(\hat{Z}=0,1|Z_\mathrm{true})$, and $L(Z_\mathrm{true}|\hat{Z}=1,1) = P(\hat{Z}=1,1|Z_\mathrm{true})$.

Each time Alice and Bob play this game, it will result in one of six sets of measurement outcomes for Bob: either his first random measurement will yield ${\hat S}_{\delta,\epsilon}=0$ and his last two $\hat{Z}$ measurements can come out one of three ways, or his first measurement will yield ${\hat S}_{\delta,\epsilon}=1$, and again his last two $\hat{Z}$ measurements can come out three ways.  
For each of these six outcomes, he will need to construct a different likelihood function and maximize it.  
He can do this readily by taking different products of the likelihood functions we just described. 
For example, if he finds ${\hat S}_{\delta,\epsilon}=0$ on his first qubit and $\hat{Z}=0$ on the other two qubits, his likelihood function for $Z_\mathrm{true}$ is:
\begin{eqnarray}
\label{eq:sampLik}
&L(Z_\mathrm{true}|{\hat S}_{\delta,\epsilon}=0)\times L(Z_\mathrm{true}|{ \hat Z}=0,0) =\\\nonumber &(\frac{1}{2}+Z_\mathrm{true}\cos\delta+\frac{1}{2}\sqrt{1-4Z_\mathrm{true}^2}\sin\delta\cos\epsilon)(\frac{1}{2} + Z_\mathrm{true})^2. 
\end{eqnarray} 
Now Bob must maximize his likelihood function (equation \ref{eq:sampLik}) over $Z_\mathrm{true}$, which depends on $\delta$ and $\epsilon$.
Since he knows he must estimate $\expec{\hat{Z}}$, and he knows what he measured (even though it was randomly chosen) he knows $\delta$, which is the angle between his measurement and ${\hat{Z}}$. 
However, since $\epsilon$ is the azimuthal angle between his measurement and the state that Alice prepares it is unknown to him (assuming that he knows nothing about Alice's state preparation).
Therefore his best strategy will be to chose it randomly, from the Haar measure.
Then he will report the value of $Z_\mathrm{true}$ at which the likelihood is maximized as his \textit{maximum-likelihood estimate} of $Z_\mathrm{true}$, $Z_\mathrm{MLE}$.

\begin{figure}
\includegraphics[scale=.5]{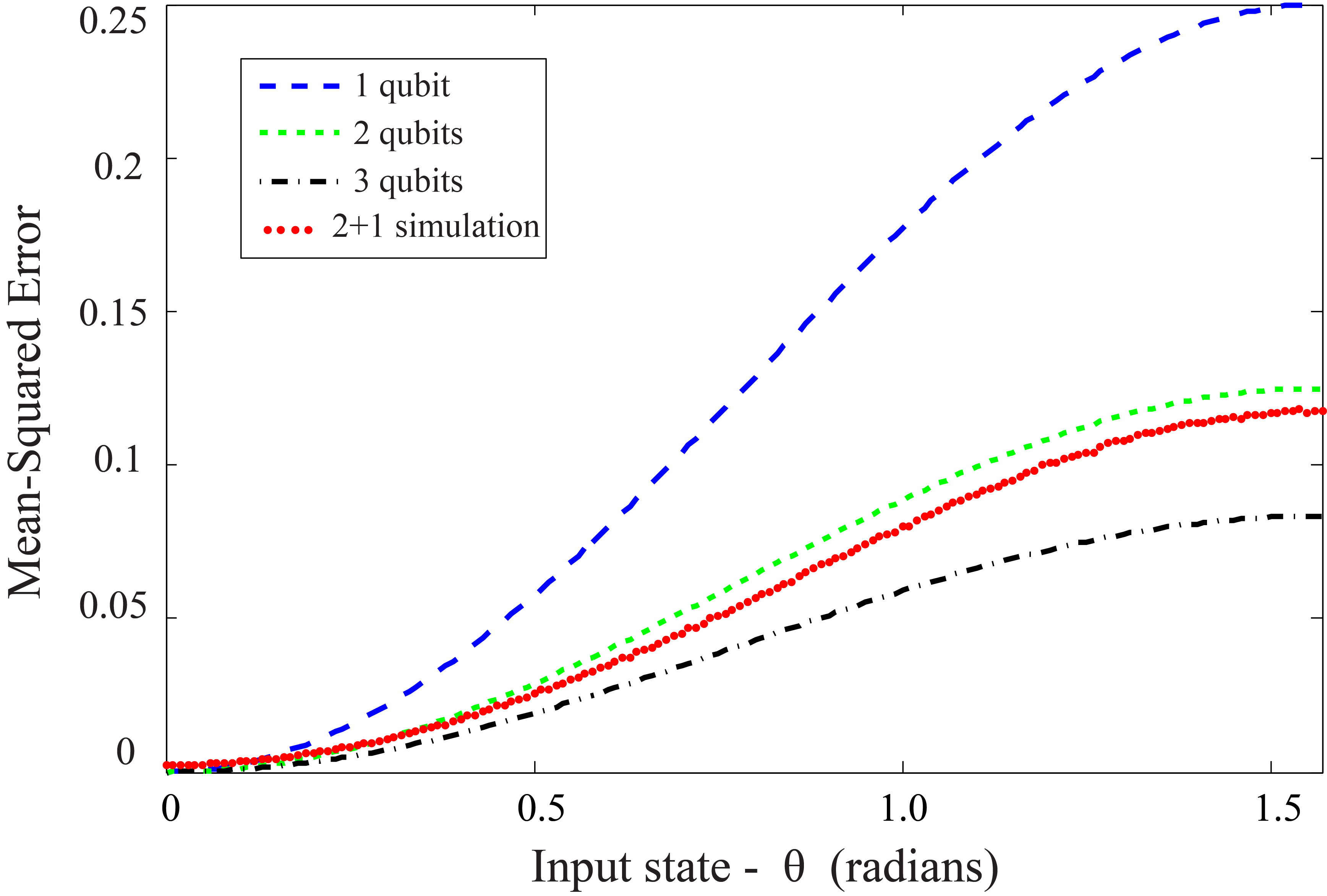}
\caption{\label{fig:5} 
\textbf{Results of Monte-Carlo Simulation ---} A plot of the mean-squared error of Bob's estimates as a function of Alice's preparation angle $\theta$, where the states are parametrized as $\cos{2\theta}\ket{0}+\sin{2\theta\ket{1}}$. 
The red data points are the simulated mean-squared difference between Bob's guess and $Z_\mathrm{true}$ when Alice sends Bob three qubits and he measures one randomly and two optimally. 
The black curve is a fit of $V_1/K$ to the simulated data, where $V_1$ is the single-qubit variance and $K$ is the only fit parameter.  
The other curves are the variances when all of the qubits are measured optimally for one (dark blue), two (light blue), and three (purple) qubits -- these are shown for comparison.  
Bob's $2+1$ scheme outperforms the two-qubit case, but does not perform as well as if all three qubits were measured optimally or as well as if he had used our compression circuit. 
}
\end{figure}

To characterize the performance of this scheme we performed a Monte-Carlo simulation. 
We averaged over $\delta$ and $\epsilon$ from the Haar measure, and observed the statistics of $Z_\mathrm{MLE}$.  
Ideally the figure of merit of $Z_\mathrm{MLE}$ would be its variance (the same figure of merit we used for our compression scheme).  
However, the average of $Z_\mathrm{MLE}$ may be different from the actual average value of ${Z}_\mathrm{true}$ (this is not true for the $Z_\mathrm{direct}$ and $Z_\mathrm{comp}$ estimators that we introduced in the main text, which do converge on ${Z}_\mathrm{true}$).  
Thus we calculate the mean-squared difference of Bob's guesses, $V_{2+1}$ (two qubits and one classical bit), from the true value of $\langle\hat{Z}\rangle$ (note that since $Z_\mathrm{direct}$ and $Z_\mathrm{comp}$ converge to $Z_\mathrm{true}$ their variance is equal to their mean-squared error). 
The result of this simulation is shown in figure 6 above (red points). 
Notice that when Alice prepares $\ket{0}$ and asks Bob to measure in $\hat{Z}$ the variance of his $2+1$ estimate is slightly larger than if he does not make use of his extra measurement.
This is because, if Bob were to simply measure two qubits in $\ket{0}$ optimally he would always find both to be spin-up, resulting in a variance of zero. 
However, his random measurement sometimes pulls his guess away from spin-up, which results in a small non-zero variance for this situation. 
The fact that the variance is larger for a specific state is not important; what matters is that Bob's variance is decreased \textit{on average} by the additional random measurement.
These simulations were also repeated for measurements of $\hat{X}$ and $\hat{Y}$, plotted as the grey dotted lines in all panels of figure 4 in the main text.

\section{Average Measurement Variance}

We can calculate the variance of all possible spin measurements, for a given state, from the variances of $\hat{X}$, $\hat{Y}$ and $\hat{Z}$.  
This is done by simply averaging them uniformly, which we will now show explicitly.

The variance of the arbitrary spin direction operator $\hat{S}_{\delta,\epsilon}$ from the previous section is:
\begin{equation}
V(\hat{S}_{\delta,\epsilon}) = \left\langle\hat{S}_{\delta,\epsilon}^2\right\rangle - \left\langle\hat{S}_{\delta,\epsilon}\right\rangle^2.
\end{equation}
Now the average variance, taken uniformly over all possible measurement directions, is:
\begin{equation}
\overline{V(\hat{S}_{\delta,\epsilon})} = \frac{1}{4\pi}\int_0^{2\pi}d{\delta}\int_0^{\pi}\sin{\epsilon} d{\epsilon} V(\hat{S}_{\delta,\epsilon}).
\end{equation}
Substituting in the forms of $V(\hat{S}_{\delta,\epsilon})$ and $\hat{S}_{\delta,\epsilon}$, the integrals can be calculated and this expression simplifies to:
\begin{eqnarray}
\overline{V(\hat{S}_{\delta,\epsilon})} = \frac{1}{3}\{  \left\langle\hat{X}^2\right\rangle + \left\langle\hat{Y}^2\right\rangle + \left\langle\hat{Z}^2\right\rangle \\ \nonumber
 - \left\langle\hat{X}\right\rangle^2 - \left\langle\hat{Y}\right\rangle^2 - \left\langle\hat{Z}\right\rangle^2\},
\end{eqnarray}
which yields the desired result:
\begin{equation}
\overline{V(\hat{S}_{\delta,\epsilon})} = \frac{1}{3}\left\{V(\hat{X})+V(\hat{Y})+V(\hat{Z})\right\}.
\end{equation}

\end{document}